\documentclass[5p,twocolumn,10pt]{elsarticle}
\usepackage{graphics}
\usepackage{amsmath,amsthm,amsfonts,amssymb}
\usepackage{amssymb}

\journal{Physica C}
\begin{document}
\begin{frontmatter}

\title{Critical current density and ac harmonic voltage generation in $\mathrm{Y}\mathrm{Ba}_{2}\mathrm{Cu}_3\mathrm{O}_{7-\delta}$ thin films by the screening technique.}

\author{Israel O. P\'erez-L\'opez}
\ead{iperez@mda.cinvestav.mx}
\author{Fidel Gamboa, V\'ictor Sosa}
\address{Applied Physics Deparment,
Cinvestav M\'erida, M\'exico, Km 6 Ant., Carretera a Progreso, A.P. 73, C.P. 97310}

\begin{abstract}
The temperature and field dependence of harmonics in voltage $V_n=V'_n-iV''_n$ using the screening technique have been measured for $\mathrm{Y}\mathrm{Ba}_{2}\mathrm{Cu}_3\mathrm{O}_{7-\delta}$ superconducting thin films. Using the Sun model we obtained the curves for the temperature-dependent critical current density $J_c(T)$. In addition, we applied the criterion proposed by Acosta et al. to compute $J_c(T)$.  Also, we made used of the empirical law $J_c \propto (1-T/T_c)^n$ as an input in our calculations to reproduce experimental harmonic generation up to the fifth harmonic. We found that most models fit well the fundamental voltage but  higher harmonics are poorly reproduced. Such behavior suggests the idea that higher harmonics contain information concerning complex processes like flux creep or thermally assisted flux flow.
\end{abstract}

\begin{keyword}
Critical current; Screening technique; flux pinning; thin films
\PACS{74.25.Ha,74.25.Qt,74.25.Sv,74.25.Nf, 74.78.Bz}
\end{keyword}

\end{frontmatter}

\section{Introduction}

In this contribution we shall determine the temperature dependence of the critical current density $J_c(T)$ for $\mathrm{Y}\mathrm{Ba}_2\mathrm{Cu}_3\mathrm{O}_{7-\delta}$ thin films by measuring the harmonics of voltage $V_n=V'_n-iV''_n$  ($n=1,3,5...$) with the screening technique \cite{classen}. The calculation is based on the Sun model \cite{sun1} and the definition of  the full penetration field $H_p$ from the critical-state model \cite{bean2} . Also an extension to higher harmonics of the this model shall be worked out and compared with experimental curves. 

\section{The Sun Model and Experimental}

We have determined the harmonics of voltage $V_n$ based on the theoretical approach developed by Sun et al. \cite{sun1}. He proposed a model to compute the ac susceptibility response of thin-film superconductors under the influence of a transversal ac magnetic field $H_a=H_{ac}\sin(\omega t)$. In response to $H_a$ shielding currents $\mathbf{J}_c$ induce a diamagnetic moment which Sun et al. found to be 
\begin{equation}
\label{momentsol}
m^{\pm}(H_a)=\pm \frac{H_pa^3}{2}\biggl[\exp{\biggl(-\frac{H_{ac}\pm H_a}{H_p}\biggr)}-\biggl(\frac{1}{2}+\eta \biggr)\biggr]
\end{equation}
where $\eta=1/2\exp(\-2H_{ac}/H_p)$ and $a$ is the radius of the film. The full penetration field stems from the Bean result, namely $H_p=J_cd/3$, where $d$ is the film thickness. Actually the quantity to be measured is the voltage $V(t)$ induced in the pick-up coil, which is conventionally measured with a lock-in amplifier and is proportional to $dm/dt$.
The voltage can be expanded as a Fourier series: $V(t)= \sum_{n=1}^{\infty}[V'_n\cos(n\omega t)+V''_n\sin(n\omega t)]$, where $V_n' =  \frac{1}{2\pi}\int_0^{2\pi}V(t)\cos(n\omega t) d(\omega t) \propto H_{ac}\omega \chi_n', $ and $V_n''  =  \frac{1}{2\pi}\int_0^{2\pi}V(t)\sin(n\omega t) d(\omega t)\propto H_{ac}\omega \chi_n''$ stands for the real and imaginary components of voltage, respectively, $\chi_n$ is the magnetic susceptibility. Here we shall deal with harmonics with $n=3,5$. The fundamental harmonic can be found elsewhere \cite{sun1}. The solutions for the real part  read:
\begin{eqnarray}
\label{sunvolresol}\footnotesize
V_3' & = &-\frac{6 m_0e^{-h}}{h}\biggl[4I_2(h)-hI_1(h)\biggr],   \\
V_5' & = & -\frac{10 m_0e^{-h}}{h^2}\biggl[48I_3(h)-12I_2(h)+h^2I_1(h)\biggr], \nonumber 
\end{eqnarray}
where $m_0=2\pi J_cda^3/3$ and $h=H_{a}/H_p$ is the reduced field, the functions $I_\nu(h)$ (with $\nu=1,2,3$) are the modified Bessel functions. The imaginary components are
\begin{eqnarray}
\label{sunvolimsol}
V_3'' & = & f(h)(3S_1-4S_2),  \nonumber \\
V_5'' & = & f(h)(5S_1-20S_2+16S_3),
\end{eqnarray}
where $f(h)=-2 m_0he^{-h}/\pi$ and the functions $S_n$ are given by
\begin{eqnarray}
\label{funcs}
S_1 & = & -\frac{2}{h^2}\biggl\{h\cosh(h)-\sinh(h)\biggr\}, \\
S_2 & = & -\frac{2}{h^2}\biggl\{h\cosh(h)-3\Bigl[S_1+\sinh(h)\Bigr]\biggr\},  \nonumber \\
S_3 & = & -\frac{2}{h^2}\biggl\{h\cosh(h)-5\sinh(h)-10S_2\biggr\}.  \nonumber
\end{eqnarray}
\begin{figure}[htp]
\begin{center}
\includegraphics[width=8cm]{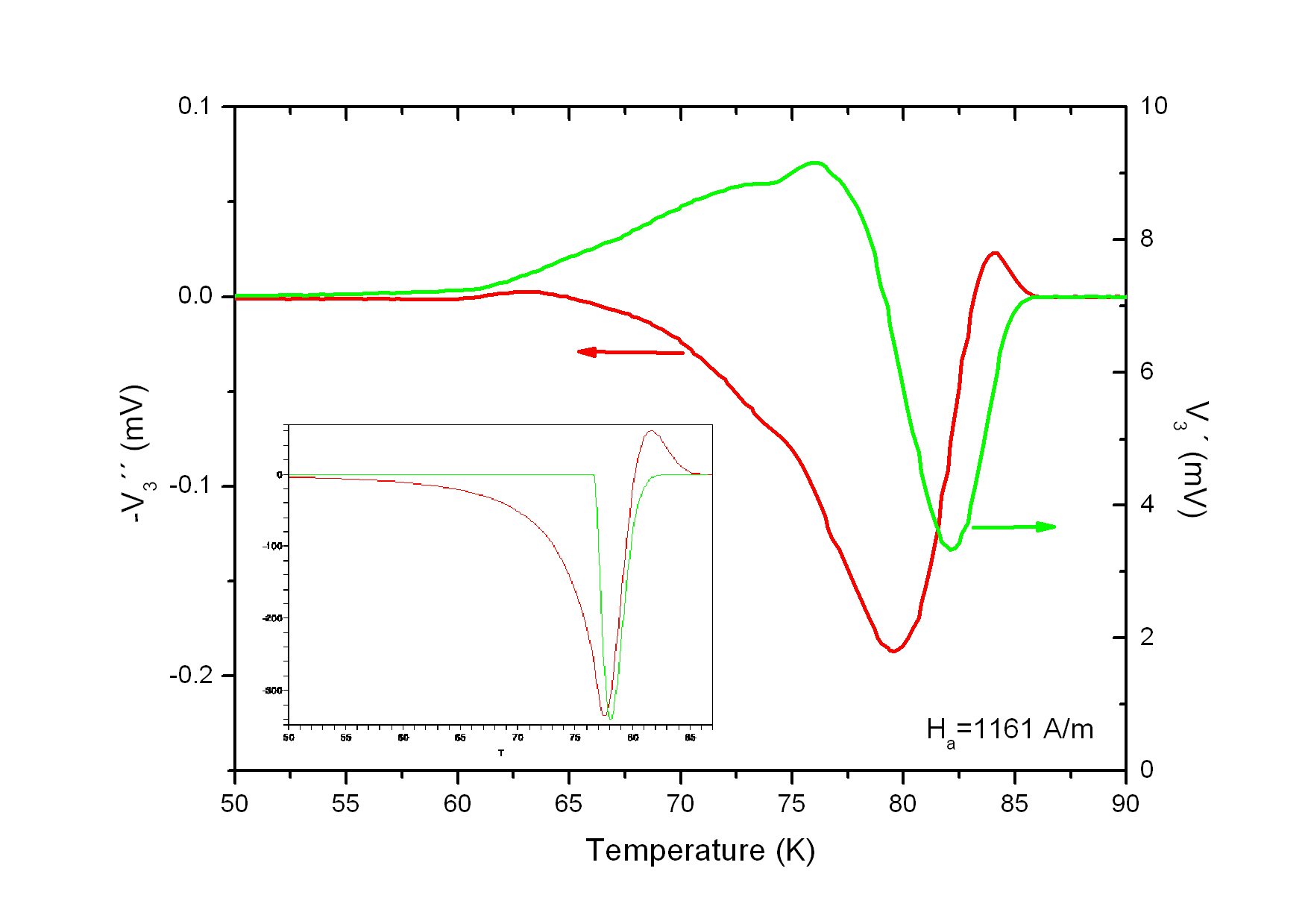}
\caption{Temperature-dependence of real and imaginary part of the third harmonic. Inset shows theoretical curves.}
\label{m23v3reim}
\end{center}
\end{figure}
To determine the critical current density $J_c(T)$ we have taken into account the criterion proposed by Acosta et al. \cite{acosta2}. He proposed that $H_p$ is the value associated with the highest peak in the ratio $\chi_3(H_{ac})/\chi_1(H_{ac})_{max}$. We also add the criterion of Xing et al. \cite{xing}, based on the measurement of the maximum of $\chi''_1(T)$. Under this method the critical current density is found to be $J_c(T_p)  =  3.157H_a/d$, where $T_p$ is the temperature at the peak. These quantities are reported in SI units. Measurements of harmonics in voltage using the screening \cite{classen} technique were carried out on superconducting thin films of $\mathrm{Y}\mathrm{B}_2\mathrm{C}_3\mathrm{O}_{7-\delta}$ with a $T_c=86.5$ K. We applied a transverse $H_{ac}$ field whose amplitude could vary from 0 to 5000 A/m with a fixed frequency of 1 KHz. For the ratio $\chi_3/\chi_1$ we have realized 11 different plots each corresponding to the temperatures: 20, 35, 45, 55, 65, 70, 75, 78, 80, 82 and 83 K. On the other hand, the temperature-dependence of the imaginary component of the fundamental harmonic  was realized for 8 different field amplitudes i.e., $H_a=$77, 118, 277, 690, 1161, 1836, 2315, 2665  A/m rms.

\section{Results and Discussion}

\begin{figure}[htp]
\begin{center}
 \includegraphics[width=8cm]{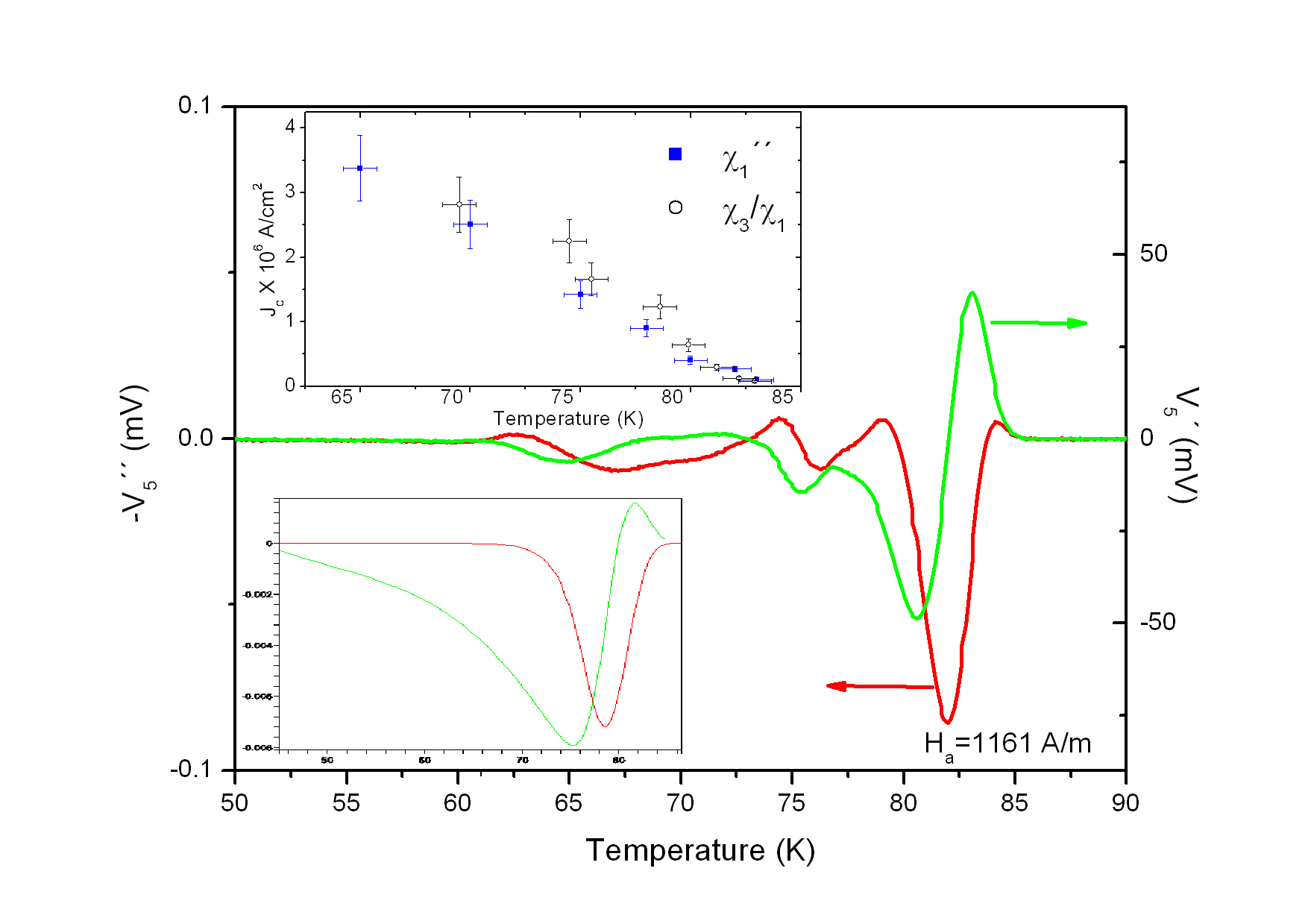}
\caption{Temperature-dependence of real and imaginary part of the fifth harmonic. Bottom inset shows theoretical curves. Top inset shows the curves for the $J_c(T)$.}
\label{m23v5reimjc}
\end{center}
\end{figure}

The upper inset in Fig. \ref{m23v5reimjc} shows the curves for the $J_c(T)$ with a systematic uncertanity of 20\%. Within the estimated uncertainties both the $\chi_3/\chi_1$ and the $\chi_1''$ approaches lay in the same order of magnitude. Therefore we conclude that in fact both approaches are equivalent.

Taking as an input, in Eqs. \ref{sunvolresol} and \ref{sunvolimsol}, the empirical law $J_c(T)=k(1-t)^n$ where $t=T/T_c$ is the reduced temperature, and $k$ and $n$ are experimental parameters we  have plotted the harmonics of voltaje. We found that in the high temperature region curves are well fitted with $n=3.4$ and $k=6.5 \times 10^{7}$ and for lower temperatures the values $n=2.8$ and $k=3.5\times 10^7$ fit well. 
In Figs. \ref{m23v3reim} and \ref{m23v5reimjc} we plot both the measured and computed curves for the third and fifth harmonics, respectively. In the real part of the third harmonic the theory can not account for  the positive peak. On the other hand, the imaginary part is qualitatively well described. The computed values for the fifth harmonic does not account for the oscillatory behavior on the left of the main peaks. As temperature goes down from above $T_c$ the main peaks agree with the theory but as the system further cools, some new features appear that suggest the presence of other mechanisms than pinning. The overall behavior for harmonics higher than the first proposes not only that pinning is the only mechanism implied in the dynamics of fluxons but also that thermally assisted flux flow (TAFF) and flux creep are present. Also a field-dependence of $J_c(B)$ is not accounted for in the model and it may be found in harmonics higher than the fundamental. However, S. Shatz et al. \cite{shatz} have shown that the Kim-Anderson model with  $J_c=J_c(B)$ also yields a universal behavior in the curves which in consequence does not produce the additional peaks. This fact has the importance conclusion: that the observed peaks must be caused only by thermal effects as the transition occurs. 

\section*{Acknowledgements}
We acknowledge the support from CONACYT.

\end{document}